\documentclass[aps,twocolumn,showpacs,superscriptaddress,citeautoscript,prl,floatfix]{revtex4-2}

\usepackage{graphicx} 
\usepackage{amsmath}
\usepackage[breaklinks, hidelinks]{hyperref}
\usepackage{amsfonts}
\usepackage{tabularx}
\usepackage{array}
\usepackage{cancel}
\usepackage{bbold} 
\usepackage{wrapfig}
\usepackage{amssymb}
\usepackage{bm}
\usepackage[T1]{fontenc}
\usepackage{textcomp}
\usepackage{enumitem}
\usepackage{esvect}
\usepackage{commath}
\usepackage[most]{tcolorbox}
\usepackage{lipsum, babel}
\usepackage{comment}
\usepackage{lineno}
\usepackage{xcolor}
\usepackage{braket}
\usepackage[capitalise]{cleveref}

\usepackage{glossaries}
\glsdisablehyper
\newacronym{imtp}{IMTP}{isolated-momentum topological phase}
\newacronym{ssh}{SSH}{Su-Schrieffer-Heeger}
\newacronym{bz}{BZ}{Brillouin zone}
\newacronym{2d}{2D}{two-dimensional}

\newcommand{\ham}{\mathcal{H}}

\let\vec\mathbf 

\hypersetup{
    colorlinks,
    linkcolor={red!80!black},
    citecolor={red!80!black},
    urlcolor={red!80!black}
}

\usepackage{adjustbox}

\begin{document}

\title{Cubic edge dispersion in a semi-Dirac Chern insulator}




\author{Marta García Olmos}
\email{mgarcia.o@usal.es}

\affiliation{Nanotechnology Group, USAL—Nanolab and IUFFyM, University of Salamanca, Plaza de la Merced, Edificio Triling\"{u}e, 37008, Salamanca, Spain
}
\affiliation{Instituto de Estructura de la Materia IEM-CSIC, Serrano 123, E-28006 Madrid, Spain}

\author{David Martín Tejedor}
\email{davidmt@usal.es}

\affiliation{Nanotechnology Group, USAL—Nanolab and IUFFyM, University of Salamanca, Plaza de la Merced, Edificio Triling\"{u}e, 37008, Salamanca, Spain}

\author{Mario Amado}
\email{mario.amado@usal.es}

\affiliation{Nanotechnology Group, USAL—Nanolab and IUFFyM, University of Salamanca, Plaza de la Merced, Edificio Triling\"{u}e, 37008, Salamanca, Spain
}

\author{Yuriko Baba}
\email{yuriko.baba@csic.es}

\affiliation{Instituto de Estructura de la Materia IEM-CSIC, Serrano 123, E-28006 Madrid, Spain}

\author{Rafael A. Molina}
\email{rafael.molina@csic.es}

\affiliation{Instituto de Estructura de la Materia IEM-CSIC, Serrano 123, E-28006 Madrid, Spain}

\date{\today}

\begin{abstract}
Topological edge states in Chern insulators are typically characterized by a linear dispersion relation inherited from the Dirac structure of the bulk Hamiltonian. Here we show that this paradigm can be fundamentally altered in systems with anisotropic semi-Dirac band structures. We introduce a minimal two-band lattice model realizing a semi-Dirac Chern insulator and determine its topological phase diagram analytically. Using a mass-domain-wall approach in a semi-infinite geometry, we derive an explicit expression for the chiral edge states and find that their low-energy dispersion scales cubically with momentum, $E(k)\propto k^3$. Numerical diagonalization of the corresponding tight-binding ribbon confirms the analytical prediction. Our results demonstrate that unconventional bulk band structures can produce qualitatively different boundary excitations, providing a route to engineering nonstandard chiral edge dynamics in topological materials and synthetic quantum systems.
\end{abstract}

\maketitle

Topological phases of matter have attracted enormous interest in condensed-matter physics due to their robust boundary modes and quantized response functions \cite{Asboth2016ShortCourse,Bernevig2013book}. In two-dimensional systems, Chern insulators represent a paradigmatic example in which the topology of the bulk bands guarantees the existence of chiral edge states traversing the bulk energy gap \cite{Haldane1988, Hatsugai1993}. These edge modes are responsible for quantized Hall transport and are protected against disorder and perturbations that do not close the bulk gap~\cite{Hasan2010,Qi2011,Anirban2023}. This relationship between bulk topology and boundary physics is captured by the bulk-boundary correspondence, which implies that the number of chiral edge modes is determined by the Chern number of the occupied bands. In most known Chern insulators, including the lattice models introduced by Qi, Wu, and Zhang~\cite{Qi2006} and the continuum models describing quantum spin Hall systems~\cite{Bernevig2006}, the edge states exhibit a linear dispersion relation near the crossing point. This linear behavior is often regarded as a generic feature of topological boundary modes.

Recent interest has focused on systems in which the low-energy band structure deviates from the conventional Dirac form. In particular, semi-Dirac systems display an anisotropic dispersion that is linear in one momentum direction but quadratic in the other. Such band structures have been predicted and observed in a variety of contexts, including engineered lattice models, optical lattices, and oxide heterostructures~\cite{Katayama2006,Dietl2008,Goerbig2008,Pardo2009,Banerjee2009,Zhong2017,Shao2024}. The interplay between this anisotropic band structure and topological band inversion leads to unconventional bulk and boundary properties that remain only partially understood \cite{Huang2015,GarciaOlmos2024,GarciaOlmos2025}. Notably, even in globally trivial phases ($C=0$), topology may emerge at isolated momenta, where a dimensional reduction leads to effective one-dimensional topological invariants (e.g., a quantized Zak phase), supporting protected modes only in restricted momentum values.

In this work we investigate the edge state of a two-dimensional Chern insulator with a semi-Dirac bulk dispersion. We introduce a minimal two-band lattice model in which the bulk bands undergo topological transitions controlled by a mass parameter and anisotropic hopping amplitudes. The topology of the model can be determined analytically by tracking the gap-closing points in momentum space, yielding regions with Chern numbers $C=0,\pm1$. Remarkably, the corresponding edge modes display a cubic dispersion near the crossing point, $E(k)\sim k^3$, in stark contrast to the usual linear behavior. In addition to the conventional chiral modes in the $|C|=1$ phases, the system hosts isolated-momentum topological states in the trivial regime. 

Using an exponential ansatz in the continuum limit, we derive an analytical expression for the edge-states dispersion in a semi-infinite system and confirm the results through numerical diagonalization of tight-binding ribbon geometries. 
Our results demonstrate that the dispersion of topological edge states can deviate significantly from the standard linear form when the bulk Hamiltonian exhibits anisotropic band touching. This provides a new route to engineering unconventional boundary excitations in topological materials and synthetic quantum systems.

\begin{figure}
     \centering
     \includegraphics[width=\linewidth]{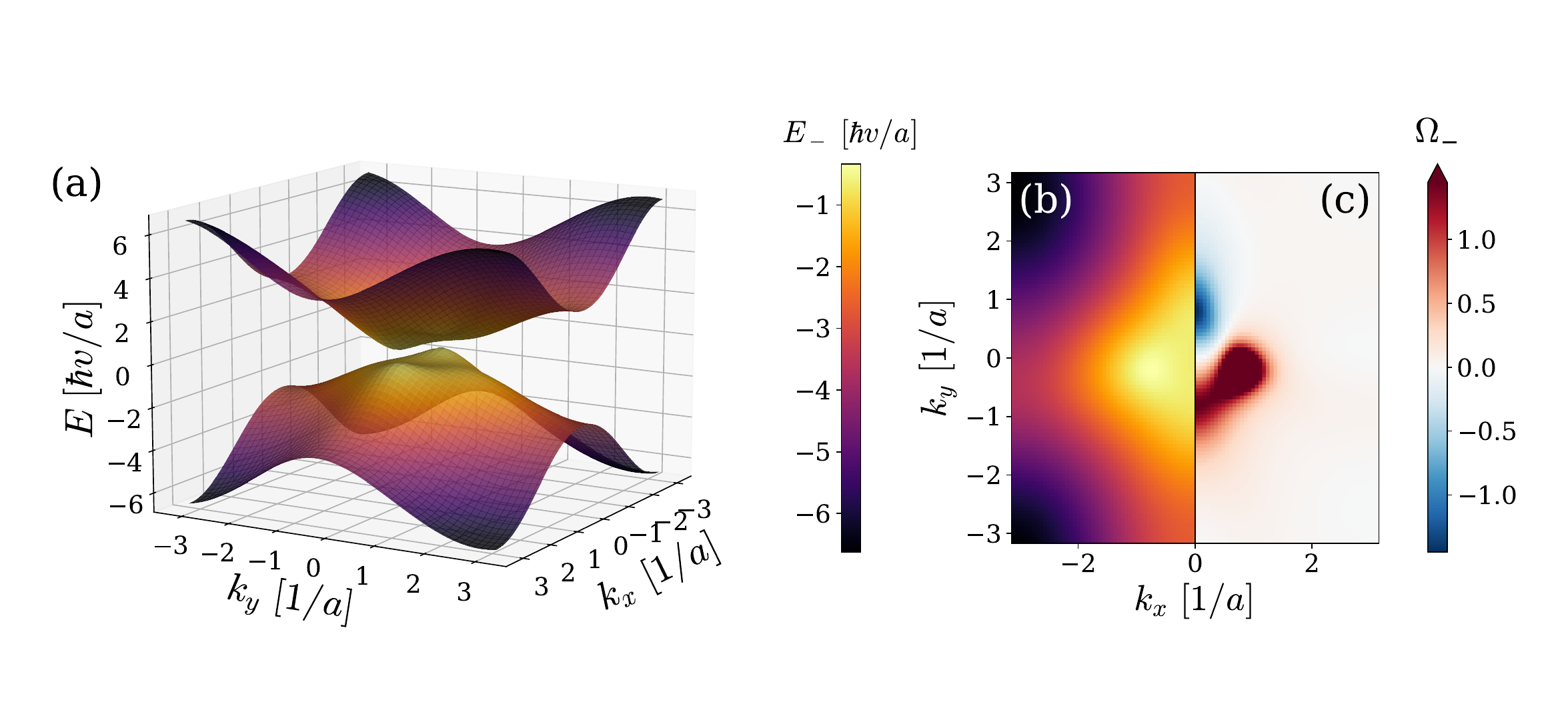}
     \caption{(a) Bulk spectrum, (b) energy $E_-$ and (c) Berry curvature $\Omega_-$ of the lower band. 
     The energy is in unit of $\hbar v /a$. The parameters are $v = 0.6$, $\alpha = 1.0$, $B_{W} = 0.5$, $B= 0.4$, $m_{\rm so} = 0.4$ and thet follow the convention of units in \cref{eq:model}.  }
     \label{fig:bulk_bands}
 \end{figure}

\emph{Model}---%
We consider a two-band lattice model defined on a square Bravais lattice, with momentum $\vec{k} = (k_x,k_y)$ measured in units of the lattice spacing $a = 1$. In the basis of a pseudospin-$\tfrac{1}{2}$ degree of freedom, the Bloch Hamiltonian is expressed as
$\mathcal{H}(\mathbf k)=\mathbf h(\mathbf k)\cdot\boldsymbol{\sigma}~$, 
where $\boldsymbol{\sigma} = (\sigma_x, \sigma_y, \sigma_z)$ is the vector of Pauli matrices and the vector components $\vec{h} = (h_x,h_y,h_z)$ are 
\begin{subequations}  \label{eq:model}
\begin{align}
    h_x(\mathbf{k}) &= v\,\sin(k_y) + 2 B\,\bigl[1-\cos(k_x)\bigr], 
    \label{eq:hx}\\ 
    h_y(\mathbf{k}) &= \alpha\,\sin(k_x)\,\sin(k_y), 
    \label{eq:hy}\\ 
    h_z(\mathbf{k}) &= m_{\rm so} \;-\; 2 B_W\!\left[\,2 - \cos(k_x) - \cos(k_y)\,\right], 
    \label{eq:hz}
\end{align}
\end{subequations}
with $v$, $B$, $\alpha$, $m_{\rm so}$ and $B_W$ real parameters with units of energy since we set $\hbar = 1$.
Near the $\Gamma$ point, $v$ plays the role of a Fermi velocity in $k_y$, while $B$ and $B_W$ generate quadratic corrections, and the parameter $m_{\rm so}$ controls the gap magnitude. $\alpha$ is a higher-order correction that controls the coupling between two momentum directions. 
The corresponding spectrum is given by $E_{\pm} (\vec{k}) = \pm |\vec{h}(\vec{k})|$ and is plotted in \cref{fig:bulk_bands} (a). The bulk bands exhibit a gapped structure with a different behavior along $k_x$ and $k_y$. This is more clearly illustrated in \cref{fig:bulk_bands} (b), where the lower band energy , $E_-(\vec{k})$, is projected in the Brillouin zone.

\emph{Analytical Chern number}---%
For the lattice Hamiltonian \eqref{eq:model}, the Chern number is obtained by integrating the Berry curvature over the Brillouin zone:
\begin{equation} 
C=\frac{1}{2\pi}\int_{\mathrm{BZ}}\Omega_-(\mathbf k)\,d^2k~,
\end{equation}
where the Berry curvature of the whole valence band is
\begin{equation} \label{eq:berry}
\Omega_-(\mathbf k)
=-\frac{1}{2}\,
\hat{\mathbf h}(\mathbf k)\cdot
\left(
\partial_{k_x} \hat{\mathbf h}(\mathbf k)\times
\partial_{k_y}\hat{\mathbf h}(\mathbf k)
\right)~,
\end{equation}
with $\hat{\mathbf h}(\vec{k}) = \vec{h}(\vec{k})/|\vec{h}(\vec{k})|$.
\cref{fig:bulk_bands} (c) represents $\Omega_-$ for the lower band displayed in panel (b) showing a clear accumulation of the Berry curvature near the regions where the energy gap is minimal. 

Since the Berry curvature is smooth whenever the spectrum is gapped, the Chern number can only change as a function of the parameters when the bulk gap closes, i.e. when   $h_x(\mathbf k)=h_y(\mathbf k)=h_z(\mathbf k)=0$. 
Due to the form of $h_y$ in Eq.~\eqref{eq:hy}, the gap-closing points must satisfy either $\sin k_y=0$ or $\sin k_x=0$ leading to two families of gaps.

\par \textit{(i) First family of gap closings.} 
For $\sin k_y=0$ and considering $h_x=0$, two candidate band-touching points are obtained for  $\vec{k} = (0, 0)$ and $\vec{k}=(0, \pi).$ From $h_z=0$ we obtain the corresponding critical parameters
\begin{equation}
m_{\rm so}=0,
\qquad
m_{\rm so}=4B_W.
\end{equation}
An additional degeneracy condition is obtained for $B = 0$, that leads to a gap closing 
at $\vec{k} = \left( \pm k_{c1}, 0 \right)$ with $k_{c1} = \arccos{\left[ 1 - m_{\rm so}/(2B_W) \right] }$ and that exists only for $0 \leq m_{\rm so} \leq 4 B_W$. 

\textit{(ii) Second family of gap closings.}
If $\sin k_x=0$, the gap closes at $\vec{k} = \left(  \pi, k_{c2} \right)$ where $k_{c2}$ satisfies $\sin(k_{c2}) = - 4B/v$ and $\cos (k_{c2}) = \left[ 3 -  m_{\rm so} / (2B_W)  \right] $. 
In this case, the gapless states are found for parameters that solve
\begin{equation} \label{eq:mso_family2}
    \left( \frac{m_{\rm so}}{2B_W} -3 \right)^2=c^2,
    \qquad
    \text{with~} c = \sqrt{1-\left( \frac{4B}{v} \right)^2}~.
\end{equation}
The gapless solution exists only if $|4B/v| \leq 1$ and $|3-m_{\rm so} / (2B_W)| \leq 1$.

\emph{Topological phase diagram}---%
Introducing the dimensionless parameter
\begin{equation} \label{eq:mu}
\mu=\frac{m_{\rm so}}{2B_W},
\end{equation}
and by considering the symmetry properties of $\Omega_-$, see End Matter \cref{eq:berry_comp}, the Chern number of the lower band can be written as
\begin{equation}
C=\operatorname{sgn}(\alpha B B_W)\,\mathcal C(\mu),
\end{equation}
where $\mathcal C(\mu)$ depends on whether the additional gap closings exist.
If $\left|{4B}/{v}\right|>1$, only the first pair of gap closings occurs, and therefore,
\begin{equation}
\mathcal C(\mu)=
\begin{cases}
1, & 0<\mu<2,\\
0, & \mu<0 \;\text{or}\; \mu>2 .
\end{cases}
\end{equation}

If $\left|{4B}/{v}\right|<1$, the additional band touchings at $k_x=\pi$ produce two further transitions, leading to
\begin{equation}
\mathcal C(\mu)=
\begin{cases}
0, & \mu<0,\\
1, & 0<\mu<2,\\
0, & 2<\mu<3-c,\\
-1, & 3-c<\mu<3+c,\\
0, & \mu>3+c .
\end{cases}
\end{equation}

At the critical values $\mu=0$, $\mu=2$, and (when present) $\mu=3\pm c$, the bulk gap closes and the Chern number is not defined.
The overall sign of $C$ is determined by the orientation of the pseudospin texture $\hat{\mathbf h}(\mathbf k)$ in momentum space, in this case by the product $\alpha B B_W$.
\begin{figure}[t]
\centering
\includegraphics[width = \linewidth]{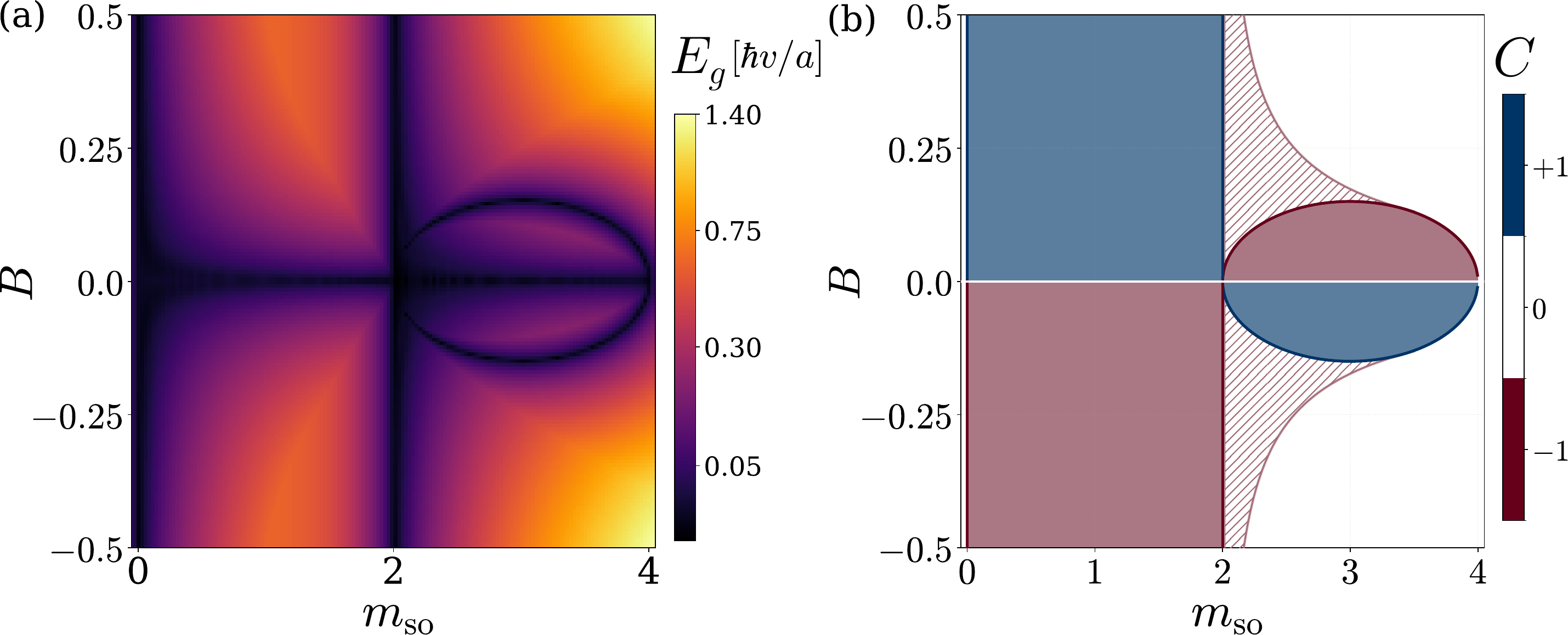}
\caption{ (a) Minimum energy gap $E_g$ 
and (b) topological phase diagram as a function of $m_{\rm so}$ and $B$. In (b), red and blue colors indicate regions with an integer Chern number $C$; the red dashed corresponds to phases with a non-trivial Zak phase along $k_y$. The remaining parameters are as in \cref{fig:bulk_bands}.
}
\label{fig:chern_phase_diagram}
\end{figure}

\par The phase diagram of the model is plotted in \cref{fig:chern_phase_diagram} as a function of the mass parameter $m_{\rm so}$ and the quadratic term $B$. Panel (a) represents the bulk gap and panel (b) the topological phases.
When $|4B/v|>1$ the spectrum closes only at $\vec{k}=(0,0)$ and $(0,\pi)$,
leading to a single topological phase with $|C|=1$.
For $|4B/v|<1$ additional gap closings occur at $k_x=\pi$,
generating two extra transitions at
the values of $m_{\rm so}$ that solves \cref{eq:mso_family2}
and an intermediate region with opposite Chern number.

\emph{Isolated-momentum topological phase}---%
Even for a trivial Chern number phase, a topological regime can emerge at isolated momentum. In this case, a system finite along $x$ host localized edge states with quadratic dispersion in $k_y$. 
The topological protection is ensured by the Zak phase~\cite{Zak1989},
\begin{equation}\label{eq:Zak}
    \mathcal{Z}_{-} (k_y) = \oint_{-\pi}^{\pi} d k_x \langle u_{-} (\boldsymbol{k}) | \partial_{k_x} | u_{-} (\boldsymbol{k}) \rangle~,
\end{equation}
where $| u_{-} (\boldsymbol{k}) \rangle$ is the Bloch state of the valence band. A non-trivial value $\mathcal{Z} = \pi$ occurs at some isolated $k_y$ points, leading to a protected topological phase only in one direction (finite in $x$), as previously studied in type I semi-Dirac systems~\cite{GarciaOlmos2024,GarciaOlmos2025}. 

The isolated-momentum topological phase can be understood via dimensional reduction, treating $k_y$ as a parameter and mapping ~\cref{eq:model} into a generalized \gls{ssh} Hamiltonian \cite{ssh_model1979},
\begin{equation}\label{eq:sshmodel}
    \mathcal{H}_{\text{SSH}} (k_x) = \left[ v + w_1 \cos k_x \right] \tau_x + w_2 \sin k_x \tau_y  
    + w_3 \tau_z~.
\end{equation}
This model describes a dimerized two-sublattice chain, where $v$ is an intracell hopping, $(w_1 +w_2)/2$ defines the first-neighbor hopping, $(w_1-w_2)/2$ is a third-neighbor hoppings and $w_3$ breaks the chiral symmetry; $\tau_{x,y,z}$ are Pauli matrices in the sublattice basis. The original \gls{ssh} model is the simplest model for a topological state in 1D and it is recovered for $w = w_i$ with $i = 1,2$ and $w_3 = 0$.

\par The Hamiltonian~\eqref{eq:model} can be written in the form of \eqref{eq:sshmodel} by defining the $k_y$-dependent terms
$V_{k_y} = v \sin k_{y} + 2B$ and
$U_{k_y} = m_{\rm so}+2B_W \left[2- \cos k_{y}\right]$, leading to 
\begin{multline} \label{eq:sshmap}
    \mathcal{H}(k_x) =  \big[ V_{k_y} \cos \theta + U_{k_y} \sin \theta + \big( 2 \sqrt{B^2 + B_W^2} \big) \cos k_x \big] \tau_x \\
    + \big[\alpha \sin k_{y} \big] \sin k_x \tau_y 
    - \left[ V_{k_y} \sin \theta - U_{k_y} \cos \theta \right] \tau_z~,
\end{multline} 
with $\tan \theta = -B_W / B$. 
In the previous expression, $\theta$ defines a rotation of the Pauli matrices basis around $\sigma_y$ that maps the semi-Dirac model $\boldsymbol{\sigma}$ to the corresponding \gls{ssh} chain $\boldsymbol{\tau}$.  

\par The gapped phases of the \gls{ssh} model are
classified in 
the ten-fold way~\cite{Chiu2016, Ryu2010, Schnyder2008} as AIII class if the chiral symmetry is imposed.
In momentum space, the chiral symmetry is translated to $\tau_z \ham (k) \tau_z = -\ham(k)$, which implies that the terms proportional to $\tau_z$ have to vanish, i.e. $w_3 = 0$. 
In this limit, the topological distinct phases are characterized by the winding number $\nu$ 
\begin{equation}
    \nu = \frac{i}{2 \pi} \int_{BZ} dk q^\dagger\partial_k q~,
\end{equation}
where $q$ is the block-off-diagonal form of the chiral symmetric Hamiltonian \cite{Asboth2016ShortCourse}. For \eqref{eq:sshmodel}, we have $q = (v + w_1 \cos(k_x) -iw_2 \sin(k_x)) /|E(k)|$ and this leads to a topological $\nu = 1$ for $|v| < |w_1|$ and trivial $\nu = 0$ for $|v|>|w_1|$.

\par Imposing chiral symmetry in~\cref{eq:sshmap} limits $m_{\rm so}$ to
\begin{equation} \label{eq:sshsym}
    m_{\mathrm{so}} \in B_W \left( 2 \pm \sqrt{ 4 + ( v/B)^2} \right)~¸
\end{equation}
while the topological $\nu = 1 $ phase requires
\begin{equation} \label{eq:sshtopo}
|v \sin k_y + 2B | < 2 |B|~.
\end{equation}
These two conditions define a parameter window in the $C = 0$ state in which the semi-Dirac model maps onto a topological \gls{ssh} chain, see more detail in the End Matter. Such region is marked in \cref{fig:chern_phase_diagram} (b) by red dashed lines.

\par
\begin{figure}[tbh]
    \centering
    \includegraphics[width=0.9\linewidth]{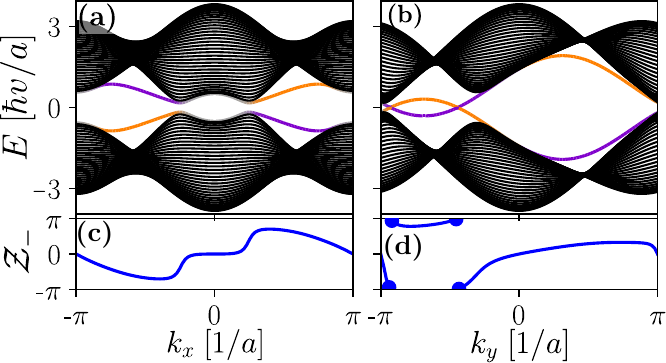}
    \caption{Band dispersion and Zak phase for semi-infinite systems in $x$ [$y$] direction for panels (a,c) [(b,d)]. In (a) $\mathcal{Z}_-(k_x)$ is obtained from \cref{eq:Zak} by replacing $x\leftrightarrow y$. 
    The parameters are as in \cref{fig:chern_phase_diagram} with $B = 0.2$ and $m_{\rm so} = 2.3$.
    }
    \label{fig:isolated_momentum_phase}
\end{figure}

 Figure \ref{fig:isolated_momentum_phase} shows the band dispersion for a semi-infinite system with translational symmetry in $x$ ($y)$-direction in the left (right) panels for a set of parameters in the isolated-momentum topological phase. Even if localized states are obtained in both systems (orange and purple bands), only in the finite system in $x$ the edge states are protected in the \gls{ssh}-sense at the isolated $k_y$ values with exactly zero energy. This can be seen using the Zak phase which varies continuously without reaching $\pi$ value as a function of $k_x$ in the trivial case in \cref{fig:isolated_momentum_phase} (c), and take exactly the value $\pi$ for the isolated $k_y$ values with zero energy in the non-trivial case in \cref{fig:isolated_momentum_phase} (d). 

\emph{Edge states in non-trivial Chern phase}---%
Next, we derive the topological edge states for the $|C|= 1$ phase in a finite system with dimensions $W\times L$. We solve the case with two boundaries located at $x=\pm L/2$ and $y= \pm W/2$. For simplicity, we treat each boundary as a semi-infinite half-plane, employing an evanescent ansatz that is valid for systems with dimensions that exceed the decay length $\lambda^{-1}$, see Supplemental Material for more details~\cite{Supp}. 

\paragraph{(i) Edge parallel to $x$.} Here 
we use the \textit{anzatz} $\psi(k_x; y) \sim e^{ik_x x}e^{\lambda (\eta y- W/2)} (\beta_{\eta}, \gamma_{\eta})^T$, with $\eta = \pm 1$ labeling the two edges   and $\text{Re}(\lambda) > 0$, to ensure proper decay into the bulk. 
The eigenvalue problem $H(k_x, -i\partial_y) \psi = E \psi$, leads to a biquadratic characteristic equation for the decay $\lambda$ with two physical solutions, $\lambda_{1,2}$ within the bulk gap. Imposing Dirichlet boundary conditions, i.e. $\psi_{\eta} (k_x; y = \eta W/2) = 0$, the solution is found to be a superposition of two modes
\begin{equation}
\psi_{\eta}(k_x; y) \propto e^{ik_x x} \left[ e^{\lambda_1 (\eta y - \frac{W}{2})} - e^{\lambda_2 (\eta y - \frac{W}{2})} \right] \begin{pmatrix} \beta \\ \gamma_{\eta} \end{pmatrix}~,
\label{eq:xdir_ansatz_main}
\end{equation}
with  $(\beta, \gamma_{\eta})^T = (\sqrt{\alpha k_x + i v}, \eta \text{sgn}(B_W) \sqrt{\alpha k_x - i v})^T$.

\par
\begin{figure}[htb]
\centering
\includegraphics[width=\linewidth]{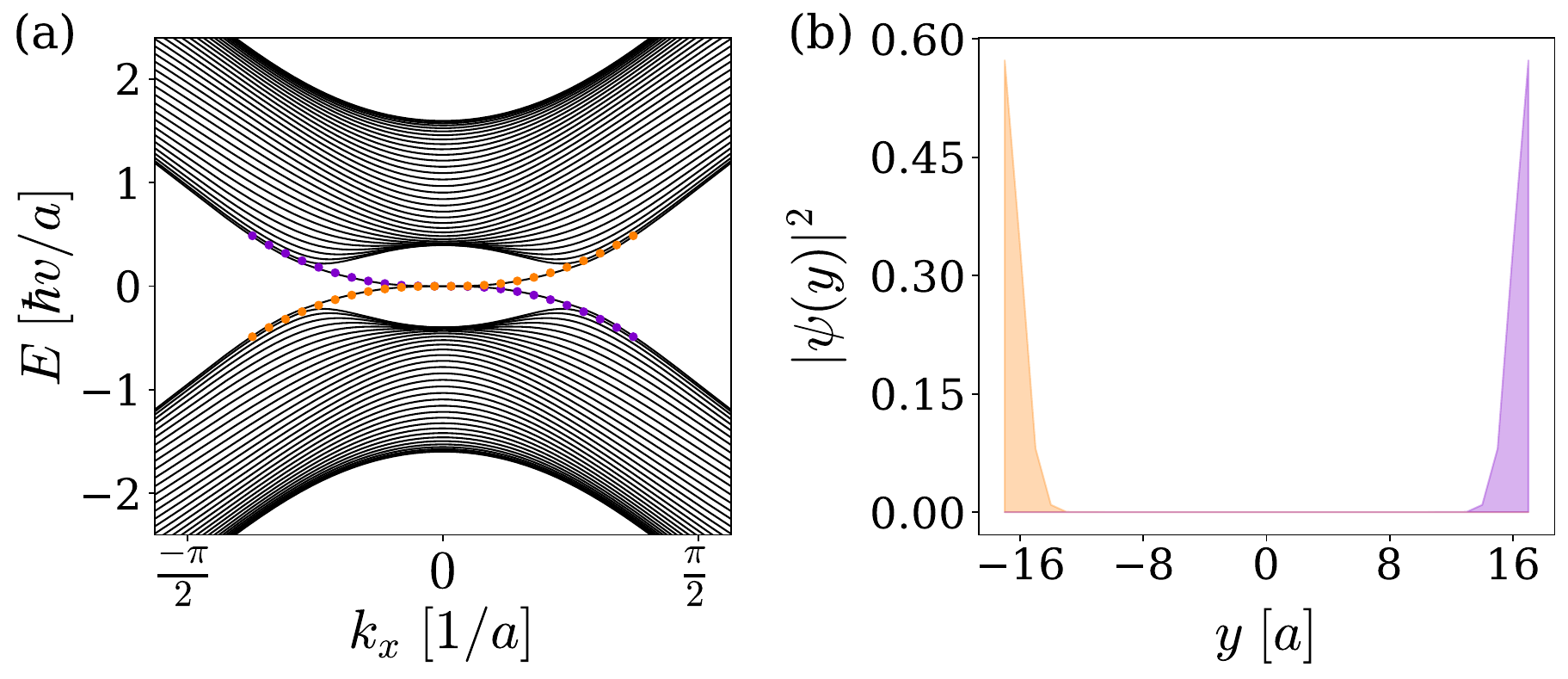}
\caption{(a) Dispersion relation  and 
(b) probability density of the edge states at $E= 0^+$ for a nanoribbon finite in the $y$-direction and $W/a = 35$. 
In (a) the analytical dispersion \eqref{eq:energy_kx} is shown in orange (purple) dotted lines for $\eta = +1$ ($\eta = -1$). The parameter values are the same as in \cref{fig:bulk_bands}.}
\label{fig:y_edges}
\end{figure}

 For these states, we obtain a dispersion relation that, near zero-energy, is cubic in momentum
\begin{equation}
    E_{\eta} = - \eta \frac{\alpha B \text{sgn}(B_W)}{\sqrt{v^2+ \alpha^2 k_x^2} }k_x^3~.
    \label{eq:energy_kx}
\end{equation}
\cref{fig:y_edges} (a) shows the excellent agreement between numerical diagonalization of a ribbon in black and the analytical solution in orange (purple) dots for upper (lower) edge. Crucially, this is a true cubic band of edge states that connects bulk valence and conduction bands.

\begin{figure}[thb]
\centering
\includegraphics[width=0.9\linewidth]{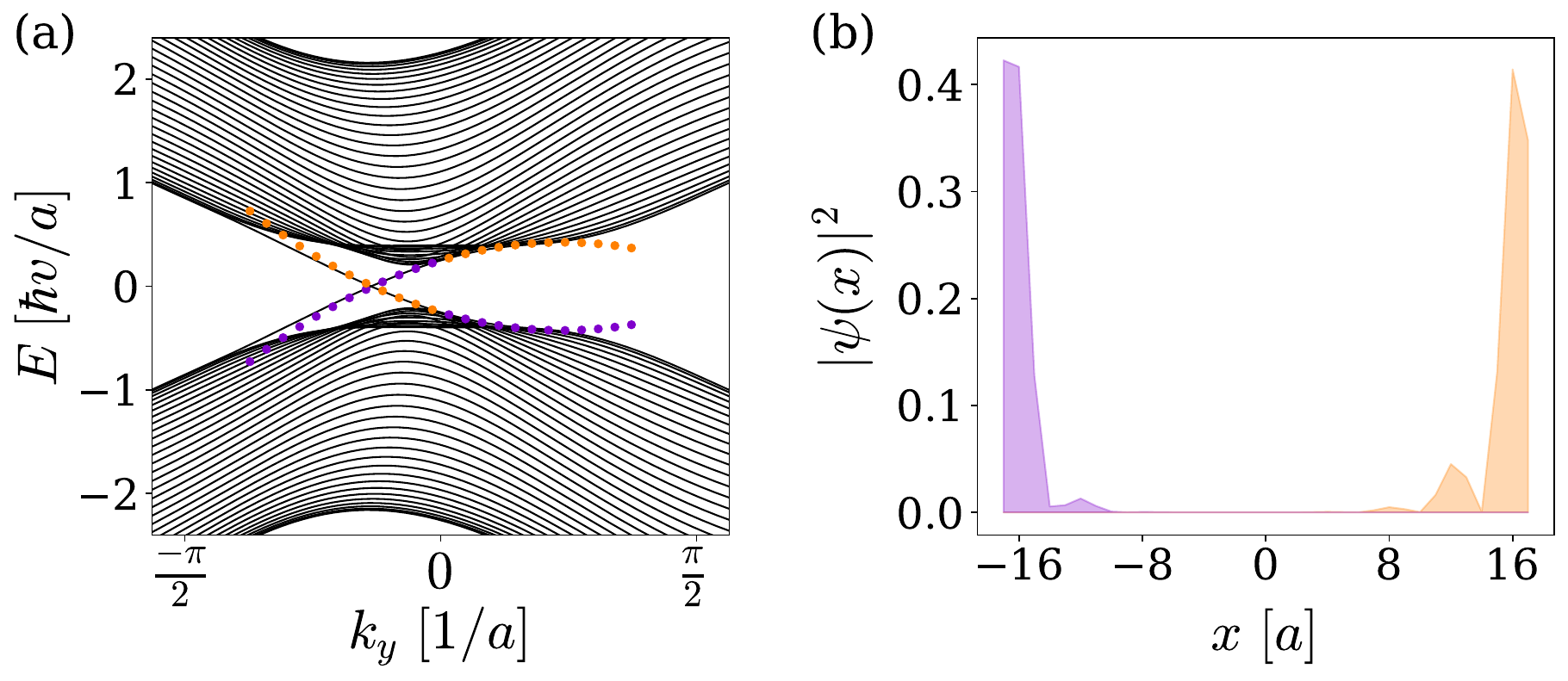}
\caption{(a) Dispersion relation  and 
(b) probability density of the edge states at $E= 0^+$ for a nanoribbon finite in the $x$ direction and with $L/a = 35$. 
In (a) the analytical dispersion \eqref{eq:energy_ky} is shown in orange (purple) dotted lines for $\eta = +1$ ($\eta = -1$). The parameter values are the same as in \cref{fig:bulk_bands}.
}
\label{fig:x_edges}
\end{figure}

\paragraph{(ii) Edge parallel to y.} For a  system with boundaries at $x = \pm L/2$, a similar procedure leads to a quartic equation for the decay lengths. The analytical dispersion of the edge states is then
\begin{equation}
    E_{\eta}(k_y) = \kappa_\eta v B_W k_y + 
     \left( m_{\rm so} - B_W k_y^2 \right) 
     \left( 1-\frac{B_W^2}{\Delta^2\kappa_\eta} \right)
     \label{eq:energy_ky}
\end{equation}
where $\kappa_\eta = 
1 - \eta \operatorname{sgn}(\alpha k_y)B /\Delta$ and $\Delta = \sqrt{B^2 + B_W^2}$. 
The spinor components are $(\beta_{\eta}, \gamma)^T = (B + \eta \text{sgn}(\alpha k_y)\Delta, B_W)^T$. 
In this case, the dispersion of the chiral states resembles the typical linear dispersion and it is in perfect agreement with the finite ribbon diagonalization in \cref{fig:x_edges}. 


Comparing the two edge directions helps elucidate the mechanism for obtaining higher-order dispersion in momentum. For the edge parallel to $y$, an independent linear term in $k_y$, i.e. in the direction tangential to the edge, is present and leads to $E\sim k_y$. In contrast, for the edge parallel to $x$, the only linear term in $k_x$ is mixed with the operator $-i \partial_y$ and it is found to vanish when projected onto the edge states. Together with the symmetry constraint on even-momentum terms, this leads to $E \sim k_x^3$, see End Matter for more details.

\emph{Conclusion}---%
In this work we have shown that a two-dimensional Chern insulator with semi-Dirac bulk dispersion can host chiral edge modes with a purely cubic dispersion with Chern number $C=\pm 1$. 
Additionally, we demonstrated that an isolated-momentum phase emerges in a parametric region within the zero Chern number regime characterized by a one-dimensional Zak phase. 

\par Our findings demonstrate that the conventional linear edge dispersion of Chern insulators can break down in anisotropic systems, giving rise to a broader class of topological boundary dynamics. 
The higher-order edge dispersions emerge in edge Hamiltonians when the momentum tangential to the edge is filtered through the normal dynamics and the leading linear term is symmetry-forbidden or projection-forbidden.
This opens a route to engineering boundary dynamics beyond the standard Dirac paradigm. In particular, the predicted cubic edge dispersion should be observable in a range of highly controllable platforms. Artificial lattices such as photonic and phononic metamaterials allow precise tuning of geometry and couplings, enabling the realization of Dirac, semi-Dirac, and higher-order band touchings, as well as topological phases with chiral edge transport~\cite{Rechtsman2013, Lu2014}. Similarly, cold atoms in optical lattices provide a versatile setting where anisotropic tunneling, artificial gauge fields, and spin–orbit coupling can be implemented~\cite{Goldman2016,Aidelsburger2018}, and where semi-Dirac points have already been engineered~\cite{Tarruell2012}. More broadly, solid-state systems with anisotropic band inversion driven by spin–orbit coupling or magnetic order offer another promising route. In all these settings, the cubic dispersion could be probed through transport, spectroscopy, or wave-packet dynamics.

The cubic dispersion implies qualitatively different dynamical properties compared to conventional Chern insulators. The group velocity scales as $v_g \sim k^2$, vanishing at the band crossing, in contrast to the finite velocity characteristic of linear chiral modes. As a result, low-energy excitations propagate slowly and the edge density of states diverges as $\rho(E)\propto |E|^{-2/3}$. In electronic realizations, this behavior may enhance the role of interactions and disorder at very low energies. In photonic and phononic metamaterials, the reduced group velocity may enable slow-light or slow-sound propagation along topological boundaries, potentially enhancing nonlinear or sensing effects. In cold-atom set-ups, the cubic dispersion could be directly probed through the dynamics of wave packets initialized at the edge: the spreading and propagation of the atomic cloud would differ markedly from the linear motion characteristic of Dirac edge states. These effects illustrate how anisotropic bulk band structures can imprint unconventional dynamical signatures on topological boundary modes.

\begin{acknowledgments}
This work has been supported by the Agencia Estatal de Investigación
from Spain (MCIN/AEI/10.13039/ 501100011033)
under Grant PID2022-136285NB-C31/C32 and
FEDER/Junta de Castilla y León Research (Grant No.
SA106P23). M G O acknowledges FEDER/Junta de
Castilla y León Research Grant No. SA121P20.
\end{acknowledgments}

%

\section*{End Matter}

\emph{Analytical Berry phase}---%
The Berry phase of the model can be written as
\begin{subequations} \label{eq:berry_comp}
\begin{multline}
    \Omega_-({\vec{k}}) = \frac{\alpha B_W}{|E_-(\vec{k})|^3}
    \left( B \omega^{\rm even} (\vec{k}) 
    + v \omega^{\rm odd} (\vec{k}) \right)
\end{multline}
with $\omega^{\rm odd (even)}({\vec{k}})$ is an odd (even) function of $k_y$: 
\begin{align}
    \omega^{\rm even}  = &   
     \left(1 -\mu \right) \sin ^2(k_x) \cos (k_y) -  \sin ^2(k_x) \cos ^2(k_y) \nonumber  \\
     &+  \left[ \cos (k_x)-1 \right] \sin ^2(k_y) \\
    \omega^{\rm odd}  = &
    -\left( 2 - \mu 
        \right) {\cos (k_x) \sin (2 k_y)}/{2}
        \nonumber \\
     & + \cos (k_x) \sin (k_y)
    + {\cos^2 (k_x) \sin (2 k_y)}/{2}~, 
\end{align}
\end{subequations}
where $\mu$ is defined in \cref{eq:mu}. Notice that all functions are even in $k_x$. Since the denominator is an absolute value, the sign of the Chern number is fixed by $\operatorname{sgn}(\alpha B B_W)$. 


\emph{SSH mapping}---%
By considering $k_y$ as a parameter and 
introducing $V_{k_y} = v \sin k_{y} + 2B$ and $U_{k_y} = m_{\rm so} + 2B_W(2 - \cos k_{y})$ we rewrite the Hamiltonian \eqref{eq:model} as 
\begin{align}
    \mathcal{H}(k_x) & = \big[-2B \sigma_x + 2 B_W \sigma_z  \big] \cos k_x\\
    & ~~~~+ \big[\alpha \sin k_{y} \big] \sin k_x \sigma_y + (V_{k_y} \sigma_x + U_{k_y} \sigma_z)~. \nonumber
\end{align}
The previous expression resembles the form of the generalized \gls{ssh} model
\eqref{eq:sshmodel} but with kinetic cosine term in the $xz$ plane of the Pauli matrices space. 
To recover the exact \gls{ssh} structure, we rotate the pseudospin space around $\sigma_y$ by an angle $\theta$, with $\tan \theta = -B_W / B$. 
This defines a new basis that maps the semi-Dirac model $\boldsymbol{\sigma}$ to the corresponding \gls{ssh} chain $\boldsymbol{\tau}$, 
\begin{equation*}
    \begin{pmatrix}
    \tau_x\\ \tau_z \end{pmatrix} 
    = R_y(\theta) 
    \begin{pmatrix}
    \sigma_x\\ \sigma_z \end{pmatrix} 
    \quad \text{with }
    R_y(\theta) = 
    \begin{pmatrix}
    \cos \theta & \sin \theta\\
    -\sin\theta & \cos \theta
\end{pmatrix}~.
\end{equation*}
This transformation leads to the Hamiltonian in \cref{eq:sshmap}. 

\par  Chiral symmetry is restored for Hamiltonian in the form of \eqref{eq:sshmodel} when $\tau_z$ term vanishes:
\begin{equation}\label{eq:em:chiralsym}
    w_3 = 0\quad \Rightarrow V_{k_y} \sin \theta + U_{k_y} \cos \theta = 0~.
\end{equation}
Substituting the definitions of $V_{k_y}, U_{k_y}$, $\theta$ and using the trigonometric identity $A \cos(x) + C \sin(x) = \sqrt{A^2 + C^2} \cos(x - \phi)$, this condition reduces to 
\begin{equation} 
    m^*_{\rm so}= 2B_W - B_W \sqrt{4 + (v/B)^2} \cos (k_y - \phi)~,
\end{equation}
where $\phi = \arctan (v/ 2 B_W)$. Since the cosine is bounded, this defines the allowed parameter window in \cref{eq:sshsym}, which are the boundaries of the red dashed region of \cref{fig:chern_phase_diagram}.

\par Sublattice symmetry restoration is a needed condition, but the system only hosts topologically protected states if $\abs{v}<\abs{w_1}$.
%
Imposing the symmetry condition \cref{eq:em:chiralsym}, we get a closed expression for the topological regime,
\begin{equation}
    |v \sin (k_y) + 2B | < 2 |B|~.
\end{equation}

\emph{Symmetry interpretation of the cubic edge dispersion}---%
In anisotropic two-dimensional topological systems, 
the conditions for a cubic or linear edge dispersion can be derived from general symmetry properties of the low-energy Hamiltonian. 
Consider an edge parallel to the coordinate $x_\parallel$, with normal coordinate $x_\perp$. Let the bulk Hamiltonian near the relevant band touching be expanded as
\begin{equation}
H(k_\parallel,k_\perp)=H_0(k_\perp)+k_\parallel V_1(k_\perp)+k_\parallel^2V_2(k_\perp)+\cdots,
\label{eq:SM_general_expansion}
\end{equation}
where $H_0(k_\perp)$ becomes an operator in real space after $k_\perp\to -i\partial_\perp$.

A \emph{linear} edge mode is expected when:
\begin{subequations}
\begin{enumerate}
\item the boundary problem for $H_0$ supports a localized bound state $\psi_0$;
\item the operator $V_1$ has a nonzero matrix element in the bound-state subspace:
\begin{equation}
\langle\psi_0|V_1|\psi_0\rangle\neq 0.
\end{equation}
\end{enumerate}
Then, the energy of the in-gap state is expanded as 
\begin{equation}
E(k_\parallel)=v_{\rm edge}k_\parallel+\cdots,
\qquad
v_{\rm edge}=\langle\psi_0|V_1|\psi_0\rangle.
\end{equation}
\end{subequations}

\par By contrast, a \emph{cubic} edge mode arises when the following three conditions hold:
\begin{subequations}
\begin{enumerate}
\item the edge Hamiltonian obeys the symmetry 
\begin{equation} \label{eq:SM_sym}
\Gamma H(k_\parallel)\Gamma=-H(-k_\parallel),
\end{equation}
for some unitary operator $\Gamma$, so that
$E(k_\parallel)=-E(-k_\parallel)$
and all even powers of $k_\parallel$ are forbidden;

\item the linear coefficient vanishes,
\begin{equation} \label{eq:SM_V1exp}
\langle\psi_0|V_1|\psi_0\rangle=0,
\end{equation}
which happens, for example, when $V_1$ is derivative-like and integration by parts kills its expectation value on a normalizable bound state;

\item no additional symmetry forces the cubic coefficient to vanish.
\end{enumerate}
Under these conditions,
\begin{equation} \label{eq:SM_cubic_disp}
E(k_\parallel)=c_3k_\parallel^3+\mathcal O(k_\parallel^5).
\end{equation}
\end{subequations}

In the present semi-Dirac model \eqref{eq:model}, \cref{eq:SM_general_expansion} reads as
\begin{subequations} \begin{align}
H_0 &= -iv\partial_y\,\sigma_x+\left(m_{\rm so}+B_W\partial_y^2\right)\sigma_z,\\
V_1 &= -i\alpha \partial_y\,\sigma_y,\\
V_2 &= B\,\sigma_x-B_W\,\sigma_z~,
\end{align}
\end{subequations}
for $x_\parallel= x$ and $x_\perp = y$. 
Notice that the anisotropic bulk Hamiltonian contains no independent linear tangential term since the only linear-in-$k_x$ is the mixed term $V_1$. 
The three conditions for cubic dispersion are the following
\begin{enumerate}
    \item the symmetry in \cref{eq:SM_sym} is given by $\Gamma=\sigma_y$ and it forces the edge dispersion to be odd in $k_x$;
    \item let $\psi_0$ be the zero-energy bound state of $H_0$. The zero mode can be chosen as an eigenstate of $\sigma_y$,  
    \begin{equation}
    \sigma_y\psi_0=s\psi_0,
    \qquad s=\pm1.
    \end{equation}    
    This way, the linear coefficient in \cref{eq:SM_V1exp} is 
    \begin{align*}
    \langle\psi_0|V_1|\psi_0\rangle
     = 
    -i\alpha s \langle\psi_0|\partial_y|\psi_0\rangle 
     = -i\alpha s \int dy\, f(y)f'(y),
    \end{align*}
    where $f(y)$ is the scalar envelope of the bound state. Since the state is normalizable,
    \begin{equation}
    \int dy\, f(y)f'(y)
    = \frac12\int dy\,\partial_y\bigl(f(y)^2\bigr)=0~,
    \end{equation}
    the linear coefficient vanishes. 
\item The exact calculation of the edge dispersion~\eqref{eq:energy_kx} shows that it has the form of \cref{eq:SM_cubic_disp} with leading cubic term 
\begin{equation}
    c_3=-\eta\,\frac{\alpha B \operatorname{sgn}(B_W)}{|v|}.
\end{equation}
\end{enumerate}

\par For comparison, consider a standard Dirac-type Chern insulator near a gap closing,
\begin{equation}
H_{\rm Dirac}(k_\parallel,k_\perp)
=
v_\parallel k_\parallel \Sigma_\parallel
+
v_\perp k_\perp \Sigma_\perp
+
m\,\Sigma_m~.
\end{equation}
Upon imposing an edge normal to $k_\perp$, one replaces $k_\perp\to -i\partial_\perp$ and obtains a Jackiw--Rebbi problem for the normal direction. The tangential momentum appears directly as an operator
$k_\parallel \Sigma_\parallel$
with a nonvanishing matrix element in the bound-state subspace. Projecting onto the edge mode yields
\begin{equation}
H_{\rm edge}(k_\parallel)\sim v_{\rm edge} k_\parallel,
\end{equation}
and hence $E(k_\parallel)\sim k_\parallel$.
This linear edge dispersion is generic whenever the bulk Hamiltonian contains an \emph{independent linear tangential coupling} that survives projection onto the bound state.

\emph{Classification viewpoint}---%
The argument above suggests a simple low-energy classification of chiral edge dynamics based on how the conserved edge momentum enters the bulk Hamiltonian near the topological band touching:


\begin{table}[h]
\footnotesize
\centering
\begin{tabular*}{\columnwidth}{c c}
\hline\hline
Bulk low-energy structure & Edge dispersion \\ \hline
Dirac-type ($k_\parallel$ enters linearly \\ and independently) & $E\sim k_\parallel$ \\ \hline
semi-Dirac-type ($k_\parallel$ enters  only \\ through mixed or higher-order terms) & $E\sim k_\parallel^3$ \\ \hline
higher-order anisotropic touching & $E\sim k_\parallel^n$ with $n>3$ possible \\
\hline\hline
\end{tabular*}
\end{table}

This should not be interpreted as a strict theorem for arbitrary models, but rather as a robust organizing principle: \emph{linear edge modes are generic for Dirac-type topological band inversion, whereas higher-order edge dispersions emerge when the tangential momentum is filtered through the normal dynamics and the leading edge velocity is symmetry-forbidden or projection-forbidden.}

In this sense, the present model provides a concrete realization of a broader mechanism by which anisotropic bulk band touching generates unconventional topological boundary dynamics.

\onecolumngrid

\setcounter{equation}{0}
\renewcommand{\theequation}{S\arabic{equation}}
\crefname{equation}{Eq.}{Eqs.}
\Crefname{equation}{Equation}{Equations}

\setcounter{figure}{0}
\renewcommand{\thefigure}{S\,\arabic{figure}}
\crefname{figure}{Fig.}{Figs.}
\Crefname{figure}{Figure}{Figures}

\section{Supplemental Material to ``Cubic edge dispersion in a semi-Dirac Chern insulator''}

\subsection{Numerical calculation of the topological invariants.}
The computation of topological invariants requires addressing the gauge freedom of Bloch eigenstates, which are defined up to a momentum-dependent phase,
\begin{equation}
    |u_n (\boldsymbol{k}) \rangle \rightarrow e^{i \theta (\boldsymbol{k})} |u_n (\boldsymbol{k}) \rangle ~.
\end{equation}

For a fixed $k_x$, the Zak phase is computed as the Berry phase accumulated along a closed loop in the $k_y$-direction. We discretize the path $k_y \in \left[ -\pi/a, \pi/a \right)$ into points $k_y^j$ for $j = 1, \dots, N$ with periodic boundary conditions $| u_n ( k_y^{N+1}) \rangle = | u_n ( k_y^1) \rangle$ [31, 32]. 
The Zak phase is obtained from the argument of the ordered product of overlaps between adjacent Bloch states, 
\begin{equation} \label{eq:ZakNumeric}
    \mathcal{Z}_n (k_x) = - \text{Im~ln}  \prod_{j=1}^{N} \left\langle u_n (k_x, k_y^j) | u_{n} (k_x, k_y^{j+1}) \right\rangle. 
\end{equation}

For the 2D bulk topology, the Chern number of the lower band is computed using the
Fukui--Hatsugai--Suzuki discretization of the Berry curvature [31]. 
The Brillouin zone is discretized into a momentum grid of points, $\boldsymbol{k}_{i,j} = (k_x^i, k_y^j)$ and a link variable $U(\boldsymbol{k}_{i,j})$ of adjacent sites in $\mu$ direction is defined as,
\begin{equation}
        U_{\mu}(\boldsymbol{k}_{i,j}) \frac{
    \langle u_n(\mathbf{k}_{i,j}) | u_n(\mathbf{k}_{i,j} + \hat{\mu}) \rangle
    }{
    \left|
    \langle u_n(\mathbf{k}_{i,j}) | u_n(\mathbf{k}_{i,j} + \hat{\mu}) \rangle
    \right|
    } ~.
\end{equation}
The Berry curvature is computed from the Wilson loop around an elementary plaquette 
\begin{equation}
    W(\boldsymbol{k}) = U_x(\boldsymbol{k}_{i,j}) U_y(\boldsymbol{k}_{i,j} + \Delta k_x) U_x^{-1}(\boldsymbol{k}_{i,j}+ \Delta k_y) U_y^{-1}(\boldsymbol{k}_{i,j})
\end{equation}
where $\Delta k_i$ denote discrete steps in momentum space. The discrete Berry flux through the plaquette is then given by
\begin{equation}
    F_{xy} (\boldsymbol{k}_{i,j}) = \text{arg} W(\boldsymbol{k}_{i,j})
\end{equation}
and global the Chern number is obtained by summing these locally gauge-invariant plaquettes' fluxes over the entire Brillouin zone, $C = (2\pi)^{-1} \sum_{\mathbf{k}_{i,j}} F_{xy}(\mathbf{k}_{i,j})$.

\begin{figure}[htb]
    \centering
    \includegraphics[width=5cm]{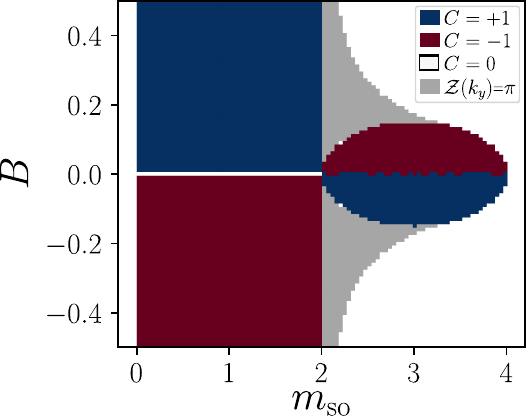}
    \caption{Numerical  topological invariants as a function of $B$ and $m_{\rm so}$. The Chern number $C$ is obtained by the Fukui-Hatsugai algorithm, while the isolated-momentum topological phase in grey is obtained by finding the parameters where the Zak phase \eqref{eq:ZakNumeric} has two roots with $\mathcal{Z}_-(k_y) = \pi$. The rest of the parameters are those given in \cref{fig:bulk_bands}.}
    \label{fig:Numerical}
\end{figure}

\subsection{Explicit calculation of the edge state dispersion}

The exact solution of the edge states that appear in a finite system of dimensions $L \times W$ is derived in this subsection. Given the system's inherent anisotropy, the edge states' dispersion depend on the boundary orientation. We therefore address edges parallel to $x$ and $y$ separately. In both cases, we employ an evanescent ansatz in semi-infinite geometries, which is a valid approximation when the system size significantly exceeds the characteristic decay length of the edge state, $\lambda^{-1}$.

\paragraph{ (i) Edge parallel to $x$.}
For an edge normal to $\hat y$, $k_x$ remains a good quantum number while the transverse momentum is replaced by the differential operator $k_y \to -i\partial_y$. We consider independent semi-infinite half planes defined as $y \leq W/2$ (upper edge) and $y \geq -W/2$ (lower edge), and explore exponentially localized solutions, $\psi_{\eta}(y; k_x) = \mathcal{N} e^{ik_x x} e^{\lambda (\eta y - W/2)} (\beta_{\eta}, \gamma_{\eta} )^T$,
that vanish exactly at the boundary. Here, $\eta = +1$ ($\eta = -1$) for the upper (lower) edge, the decay length satisfies Re$(\lambda) > 0$ ensures decay into the bulk and $\mathcal{N}$ is a normalization constant. The Hamiltonian reads
\begin{align}
H (k_x, -i \partial_y)  =
\bigl(-iv\partial_y + Bk_x^2\bigr)\sigma_x -i\alpha k_x \partial_y\,\sigma_y 
 +\bigl(m_{\rm so}-B_W k_x^2 + B_W\partial_y^2\bigr)\sigma_z ~.
\end{align}
The eigenvalue problem, $H (k_x, -i \partial_y) \psi_{\eta} = E \psi_{\eta} $ leads to a quartic equation in $\lambda$, whose coefficients satisfy that the cubic term vanishes and the linear term is purely imaginary. From Cardano-Vieta relations, the roots appear in pairs $(\lambda_i, -\lambda_i^*)$ with $i = 1,2$, and then, only two of them satisfy the physical requirement $\text{Re}(\lambda) > 0$. Imposing Dirichlet boundary conditions $\psi_{\eta = \pm} (y = \pm W/2; k_x) = 0$, we get 
\begin{equation}
    \psi_{\eta} = \mathcal{N} e^{i k_x x} \left[  e^{\lambda_1 (\eta y-W/2) } -  e^{\lambda_2 (\eta y-W/2) }\right] \begin{pmatrix} \beta_{\eta} \\ \gamma_{\eta}  \end{pmatrix} ~.
    \label{eq:xdir_ansatz}
\end{equation}
Substituting this ansatz into the eigenvalue equation provides the energy in terms of the decay lengths $\lambda_{1,2}$,
\begin{equation}
    E = m_{\rm so} - B_{W}k_x^2 - B_W(\lambda_1\lambda_2) \mp \frac{B_W B k_x^2}{\alpha k_x + i v}(\lambda_1+ \lambda_2)~,
\end{equation}
where the sum and the product of the decay lengths can be obtained from $H (k_x, -i \partial_y) \psi_\eta (y; k_x) \big|_{y =\eta W/2} = 0$,
\begin{subequations} \label{eq:xdir-lambda}
    \begin{align} 
    &(\lambda_1 + \lambda_2)^2 = \frac{v^2 + \alpha^2 k_x^2}{B_W^2} ~, \\
    &\lambda_1 \lambda_2 = \frac{m_{\rm so} - B_W k_x^2}{B_W} + \eta \frac{i v B k_x^2}{|B_W|\sqrt{v^2 + \alpha^2 k_x^2}} ~.
\end{align}
\end{subequations}
Finally, the dispersion relation in terms of the momentum is given by,
\begin{equation}
    E_{\eta} = - \eta \frac{\alpha B \text{sgn}(B_W)}{\sqrt{v^2+ \alpha^2 k_x^2} }k_x^3 ~,
\end{equation}
and the spinor components are, $(\beta, \gamma_{\eta})^T = (\sqrt{\alpha k_x + i v}, \eta \text{sgn}(B_W) \sqrt{\alpha k_x - i v})^T$. 

\paragraph{(ii) Edge parallel to $y$.} For edges normal to $x$, $k_y$ remains a good quantum number but $k_x \rightarrow -i \partial_x$. Again, we consider isolated edges, treating them as semi-infinite half-planes defined by $x \leq -L/2$ (left edge) and $x \leq L/2$ (right edge) and use the ansatz, $\psi_{\eta} (x;k_y) = \mathcal{N} e^{ik_y y} e^{\lambda ( \eta x - L/2)} (\beta_{\eta}, \gamma_{\eta})^T$, with $\eta = +1$ ($\eta = -1$), for the left (right) boundary. The Hamiltonian becomes, 
\begin{align}
    H(-i \partial_x, k_y) = (-B \partial_x^2 + v k_y) \sigma_x - i \alpha k_y \partial_x \sigma_y  +  (m_{\rm so} + B_W \partial_x^2 - B_W k_y^2)\sigma_z ~.
\end{align}
In this case the eigenvalue problem, $H (-i\partial_x, k_y) \psi_{\eta}= E \psi_{\eta}$, yields to a biquartic equation in $\lambda$. For each energy there are four solutions, but only two satisfy the physical requirement to be boundary solutions. Considering this and the boundary condition $\psi_{\eta}(x = \pm L/2; k_y) = 0$ we get the analogous expression of the Eq. \eqref{eq:xdir_ansatz} for this edge,
\begin{equation}
    \psi = \mathcal{N} e^{i k_y y} \left[  e^{\lambda_1 (x-L/2) } -  e^{\lambda_2 (x-L/2) }\right] \begin{pmatrix} \beta_{\eta} \\ \gamma_{\eta}  \end{pmatrix} ~.
    \label{eq:ydir_ansatz}
\end{equation}
Introducing $\Sigma_{\eta} (k_y) = B(\lambda_1 + \lambda_2) + \eta \alpha k_y$
the energy reads,
\begin{align}
    E_{\eta} =  \frac{1}{\Sigma_{\eta} (k_y)} \big( B  \left[m_{\rm so}-B_W k_y^2 + v k_y B_W\right] (\lambda_1+ \lambda_2) 
    - \eta \big[ \alpha k_y B_W \lambda_1 \lambda_2 + \alpha k_y (m_{\rm so}-B_W k_y^2) \big] \big) ~.
\end{align}
where
\begin{equation}
    (\lambda_1+ \lambda_2)^2 = \frac{\alpha^2 k_y^2}{B^2 + B_W^2} ~, 
    \quad
    \lambda_1 \lambda_2 = \frac{m_{\rm so}B_W - B_W^2 k_y^2 - B v k_y}{B^2+ B_W^2}~.
\end{equation}
Substituting these expressions and making some algebra, we arrive to the dispersion reported in the main text,
\begin{equation}
    E_{\eta}(k_y) = 
     \left( m_{\rm so} - B_W k_y^2 \right) 
     \left( 1-\frac{B_W^2}{\Delta^2\kappa_\eta} \right)
     + \kappa_\eta v B_W k_y  
\end{equation}
where $\kappa_\eta = 
1 - \eta \operatorname{sgn}(\alpha k_y)B /\Delta$ and $\Delta = \sqrt{B^2 + B_W^2}$.  The spinor components, not normalized, are $(\beta_{\eta}, \gamma)^T = (B + \eta \text{sgn}(\alpha k_y)\Delta, B_W)^T$. 

\end{document}